\documentclass{aa}
\usepackage{graphics}

\begin{document}

\thesaurus{08(02.04.2, 08.08.01, 08.12.1)}

\title{Microscopic diffusion and subdwarfs}

\author{P. Morel\inst{1} \and A. Baglin\inst{2} }

\institute{D\'epartement Cassini, UMR CNRS 6529, Observatoire de la C\^ote 
d'Azur, BP 4229, 06304 Nice CEDEX 4, France
\and 
DESPA, URA CNRS 264, Observatoire de Paris-Meudon, 92195 Meudon
Principal CEDEX, France.
}

\offprints{P. Morel}
\mail{Pierre.Morel@obs-nice.fr}

\date{Received date / Accepted date}

\maketitle

\begin{abstract}
The recent distance determinations by HIPPARCOS have allowed
to locate very precisely a rea\-son\-able set of subdwarfs in the HR diagram.
A detailed comparison with evolutionary models is now possible and it
has been recently claimed that the observed 
lower part of the main-sequence is cooler than the computed one.
In this paper we show that microscopic diffusion could be the physical
explanation of such a systematic shift.
We emphazise the fact that, as diffusion alters surface abundances, the initial metal
content of the stellar material is higher than the present one. A ``calibration" procedure
is necessary
 when comparing observed and theoretical HR diagram, to take into account this effect.

\end{abstract}

\section{Introduction}\label{sec:int}
Microscopic diffusion is a basic
phenomenon, but, in the physical conditions of stellar interiors
the small diffusion velocities  imply long time
scales to obtain significant
mo\-di\-fi\-ca\-tions on the global parameters of real stars. To be efficient,
the medium has to be quiet enough,
so that large scale motion cannot prevent the settling.
Acting essentially in radiative zones, it is now been advocated 
to explain a variety of observed facts, as lithium abundances in young and old 
population (Vauclair \& Charbonnel \cite{sv}), abundances anomalies
in different classes
of young stars (Hui Bon Hao \& Alecian \cite{ha} and references herein),
glo\-bu\-lar cluster evolution and
age determination (Chaboyer et al. \cite{cha}).
In the Sun, the insertion of microscopic diffusion in the modeling improves
significantly the agreement between the theoretical
 and the "seismic" model (see e.g.
Christensen-Dalsgaard et al. \cite{daa}; Richard et al. \cite{ric};
Morel et al. \cite{mpb};
Brun et al. \cite{btm}). In the present state of art
the computed sound speed, beneath the solar convection zone,
agrees within a rms
discrepancy better than 0.2\% (Gough et al. \cite{get});
the predicted values of the radius at the bottom
of the convection zone and the helium abundance at the solar surface 
  agree within the error bar with their values
inferred from helioseismology (see e.g. Basu \cite{b}).

Subdwarfs represent a clue in the general framework of stellar evolution
in particular to estimate 
the globular clusters ages. A precise knowledge of their internal
structure and of the
stage of evolution is then badly needed. 

 As in such old objects the evolutionary time scale becomes
of the same order as the diffusion one, significant effects
due to this physical process
 are expected on their global properties. 

Several papers have already mentioned the influence of microscopic
diffusion in old objects (see e.g. Chaboyer et al. \cite{cha}; Mazitelli
et al. \cite{maz}; Castellani et al. \cite{cas}), but focussed
generally on the consequences on the age  of the oldest globular clusters.

Up to now, the poor quality of the observed parameters of population II objects
did not allow to refine their theory.
The recent distance determinations of subdwarfs by HIPPARCOS
(Perryman et al. \cite{pera}) has revolutionized the field.
Combining these results with large improvement of effective temperature and
bolometric corrections for low metal atmospheres (Alonso et al. \cite{alo}),
a reasonable set of subdwarfs have been located very precisely  in
the HR diagram. A detailed comparison
with evolutionary
models is now possible and requires the same precision
in the description of the physics of their interiors.

It has been recently emphasized that difficulties arise when trying
to fit observed and computed main-sequences, as models look hotter than
the real objects 
(Baglin \cite{ba}; Lebreton et al. \cite{lpf}; Cayrel et al. \cite{rc};
Lebreton et al. \cite{leb}).

Several effects can been advocated to explain this discrepancy,
as for instance, 
the treatment of the superadiabatic outer layers or observational
bias due to NLTE effects,
in the effective temperature scale and in the abundance determinations
(Thevenin \& Idiart \cite{ti99}).

We propose in this paper to document
the role of microscopic diffusion in these stages of evolution,
and to quantify its
influence on the global parameters of
these stars and on their position in the HR diagram.

It is well known that, on the main-sequence, variations of helium $Y$ and of
the metal content $Z$
act in opposite directions~: while decreasing $Y$ reduces the effective
temperature $T_{\rm eff}$ and the 
luminosity $L$, a decrease of the metal content increases both $L$ and
$T_{\rm eff}$.
Microscopic diffusion, which acts on both $Y$ and $Z$, 
 creates a 
stratification of chemicals in radiative zones.
The global effects on the observables i.e. the modification of the HR
diagram position, but
also the changes of the  surface metal abundance, are quite subtle;
they have to be
taken into account altogether 
to perform a precise comparison with observations.

 The paper is organized as follows~: the set
of stellar models is listed in
Sect.~\ref{sec:mod}, whereas the input physics is described in Sect.~\ref{sec:phy}.
The effects of microscopic diffusion on the evolution of
subdwarfs are presented in
Sect.~\ref{sec:diff}, the  "calibration" procedure needed when comparing
theoretical predictions and observations is developped in
Sect.~\ref{sec:cal} , whereas the influence on
the HR diagram and on the shape of the isochrones
as a function of their present observed metal content is described
in Sect.~\ref{sec:HR}.
Future prospects and possible tests of this hypothesis
are suggested in Sect.~\ref{sec:fut}.

\section{The stellar models}\label{sec:mod}

As subdwarfs belong to population II, their initial
helium content is close to the primeval value.
Then stellar models depend only on age and metal content,
assuming that the physical description is completely settled.

In old metal poor stars, $\alpha$ elements enrichment seems
to be always present
(see e.g. Wheeler et al. \cite{wh}), though large variations exist from one
star to the other. We have used in all our models the
mixture of Allard (\cite{all}) with the $\alpha$-enrichment
relative to the Sun, $\mathrm{[\alpha/Fe]=0.3}$.
Four initial metal contents, represented by the standard
 [Fe/H] observable, have been considered~:  
$\mathrm{[Fe/H]}\sim -0.75,\ -0.94,\ -1.24\ {\rm  and}\ -1.72$. 
The initial total amount of helium, i.e.
\element[][3]{He}\,+\,\element[][4]{He}, per
mass is fixed to its primordial value assumed to
be $\rm{Y=0.24}$ (Izotov et al. \cite{iz}). With respectively
$\mathrm{Z_{Fe}}$, X and Z as
the iron, hydrogen and heavy elements mass ratio, we use the relation~:
\begin{equation}\label{eq:fesh}
\mathrm{[Fe/H]\equiv\log(Z_{Fe}/Z) + \log(Z/X) - \log(Z_{Fe}/X)_\odot}
\end{equation} 
to relate the observed metallicity and metal content. Here
$\mathrm{\log(Z_{Fe}/X)_\odot\equiv-2.753}$ for the solar
mixture of Grevesse \& Noels, (\cite{gn}), and 
$\mathrm{\log(Z_{Fe}/Z)\equiv-1.371}$ for the Allard's
$\alpha$-enriched mixture. 

The initial mass ratio X and Z are
derived from the initial values of Y and [Fe/H] as given in Table~\ref{tab:x}.

\begin{table}
\caption[]{[Fe/H] and 
initial abundances, per mass unit, of hydrogen X and heavy elements Z.
}\label{tab:x}
\begin{tabular}{cccc} \\  \hline \\
Labels &[Fe/H] &  X & Z   \\ \\ \hline \\
                  
07     &-0.75&0.7544&0.0056\\
09     &-0.94&0.7564&0.0036\\
12     &-1.24&0.7582&0.0018\\
17     &-1.72&0.7594&0.0006\\ \\ \hline \\
\end{tabular}
\end{table}

To describe the lower part of the main-sequence we present
here models of masses varying from
$0.6$ to $0.85 M_\odot$ evolving up to $10$\,Gy. Below $0.6 M_\odot$,
evolution is extremely slow,
and above $0.85 M_\odot$ subdwarfs have left the main-sequence.
 As during the rapid  pre main-sequence
phase, the stellar models are almost fully mixed by convection, 
the microscopic diffusion is inefficient, and starting the computations
at the homogeneous main-sequence stage is sufficient for our purpose.

\section{Physics of stellar models}\label{sec:phy}
\paragraph{Nuclear network and chemical mixture.}
The general nuclear network we used contains the following species~:
\element[][1]{H},
\element[][3]{He},
\element[][4]{He},
\element[][12]{C},
\element[][13]{C},
\element[][14]{N},
\element[][15]{N},
\element[][16]{O},
\element[][17]{O},
and Ex. Ex is a fictitious
mean non-CNO heavy element with atomic mass 28 and charge 13, which
complements the mixture.
With respect to time,
due to the diffusion processes, the
abundances of heavy elements are enhanced toward the center;  
Ex  mimicks that enhancement for the non CNO metals 
which contribute to changes of Z, then to opacity variations.
We have taken into account the 
most important nuclear reactions of PP+CNO cycles (Clayton, \cite {cy})
with the species $^2$H, $^7$Li, $^7$Be at equilibrium.
The relevant nuclear reaction rates
are taken from Caughlan and Fowler (\cite{cf}); weak screening is assumed.

For helium the initial isotopic ratio is fixed at
\element[][3]{He}/\element[][4]{He}$=4.19\,10^{-4}$ (Gautier \& Morel \cite{gm}).
In the mixture of heavy elements, the ratios between the CNO species are
set to their $\alpha$-enriched Allard (\cite{all})
values (in number) C~: 0.147909 , N~: 0.038904 and O~: 0.616594, 
with the isotopic ratios~:
\element[][13]{C}/\element[][12]{C}$=1.11\,10^{-3}$,
\element[][15]{N}/\element[][14]{N}$=4.25\,10^{-3}$,
\element[][17]{O}/\element[][16]{O}$=3.81\,10^{-4}$
(Anders \& Grevesse, \cite{ag}).

\begin{figure}
\resizebox{7cm}{!}{\includegraphics{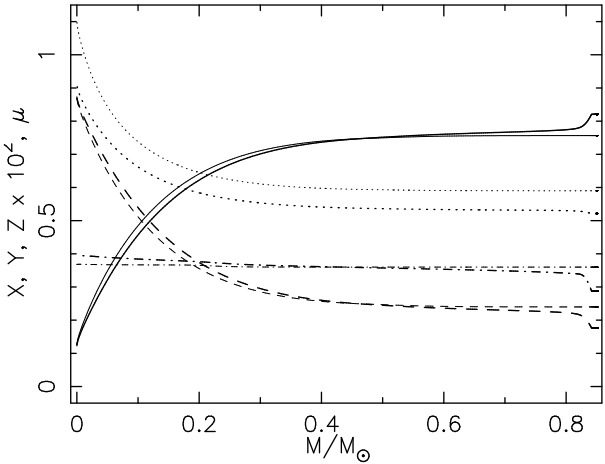}}
\caption{
Profiles of abundances, in mass, of hydrogen $X$ (full), helium $Y$ (dash),
heavy elements $Z$ (dot-dash) and mean molecular weight $\mu$ (dot)
for two $0.85M_\odot$ models evolved up to 10\,Gy, respectively
with (heavy) and without (thin) microscopic diffusion. 
}\label{fig:abon}
\end{figure}

\paragraph{Diffusion.}
Different processes are participating to element separation or mixing.
Whereas gravitational settling creates stratification, 
hydrodynamical instabilities can, 
at least in some phases of evolution, generate  macroscopic motions
which tend to reduce the chemical inhomogeneities.
Presently, no general description exists of these
processes, which could be easily incorporated in  stellar evolution
calculations.
 As the purpose of this paper is to
evaluate the influence of elements segregation in comparing theory and
observations of subdwarfs, we did  not take into account any of these processes. 
Let us note that rotation, which has been identified as one major
cause of partial mixing is
almost absent in these old low mass objects.

Microscopic diffusion is described using the formalism of 
Michaud \& Proffitt (\cite{mipr}) valid for main-sequence stars.
The radiative forces are not taken into account. For $Y$
the mass conservation equation gives~: $V_Y=-\mathrm{X} V_X / \mathrm{Y}$
where  the original equation has been slightly modified
 to take into account the heavy elements; 
$V_X$ and $V_Y$  are respectively the diffusion velocities of
hydrogen and helium.
This description is parameter free, depending only on known coefficients.
A more complete discussion of the method used to treat diffusion
is given in Morel et al. (\cite{mpb}).

\paragraph{Equation of state and Opacities.}\label{sec:eos}
We have used the EFF equation of state (Eggleton et al. \cite{eff})
sufficient for our purpose
in this mass range.

We have chosen the  Livermore Library (Iglesias \& Rogers \cite{ir}) 
with the $\alpha$-enriched mixture of Allard (\cite{all}), complemented at 
low temperature opacities by the
Kurucz's (\cite{kur}) $\alpha$-enriched tables. 
Unfortunately there is no available common mixture in these
libraries; we have retained the closest ones with a rough smoothing in the
extreme low temperature atmospheric layers. 

The opacities and equation of state are functions of
the heavy elements content Z, through the number of free 
electrons and the abundances of efficient absorbers which do not necessarily
belong to the nuclear network e.g. $^{56}$Fe.
Due to diffusion and nuclear reactions, Z changes
as the star evolves, as well as the ratios between the abundances of chemicals.
In the calculation of opacities and equation of state, 
$Z$ is separated in two parts~: the first one consists of
the chemicals heavier than helium which, belonging to the CNO nuclear network,
are both diffused and nuclearly processed; Ex,
the second part, is only diffused. Hence in the estimate of Z, 
the changes of CNO abundances
caused by diffusion, nuclear processes and the
effects of the gravitational settling of the  heaviest non-CNO
species are taken into account.

\paragraph{Convection.}
In the convection zones the temperature gradient is
computed according to the standard mixing-length theory, with
the mixing-length defined as $l\equiv \lambda H_{\rm p}$,
where $H_{\rm p}$ is the classical pressure scale height.
The constancy of $\lambda$ has been demonstrated
for population I models close to the main-sequence (see e.g. Fernandes
et al. \cite{fe}), and calibrations of $\lambda$ from 2D simulations
(Ludwig et al. \cite{lud}) lead to
approximately the same result; so,  we assume that
$\lambda$ is constant and equal to the value obtained for a solar model
using the same physics, namely $\lambda\equiv1.7$. 
This hypothesis has no influence on the results of this paper devoted
to an estimate of the differential effect of microscopic diffusion.

In the models with diffusion the convection zones are mixed via
a strong turbulent diffusion  coefficient, which produces an homogeneous
composition.

\paragraph{Atmosphere.}
An atmosphere is restored using the Hopf's $T(\tau)$ law ~:
$T^4=\frac 34 T_{\rm eff}^4 [\tau+q(\tau)]$ (Mihalas \cite{ml})
with $\tau$ as the Rosseland optical depth.
The connection with the envelope is made at the optical
depth $\tau_{\rm b} = 10$, where the diffusion approximation for radiative
transfer becomes valid. In the convective part of the atmosphere,
a numerical trick
(Henyey et al. \cite{hvb}) is employed in connection with the purely
radiative Hopf's law in order to ensure the continuity of gradients
at the limit between the atmosphere and the envelope. 
At each time step, the radius $R_\star$ of the model
is taken at the optical depth $\tau_\star\simeq 0.6454$ where
$T(\tau_\star)=T_{\rm eff}$;
the mass of the star $M_\star$, is the mass inside the sphere of
radius $R_\star$.
The external boundary is fixed at the optical
depth $\tau_{\rm ext}=10^{-4}$, where a boundary condition on the
pressure is expressed as~:
$P(\tau_{\rm ext})=g\ /\ \kappa\tau_{\rm ext}$, with $g$ as the gravity
and $\kappa$ as the Rosseland mean opacity.

\paragraph{Numerics.}

The models have been computed using the CESAM code (Morel \cite{mp96}).
Typically each evolutionary track needs of the order of 85 models of about
500 mass shells.
The accuracy of the numerical scheme is one for the time
and three for the space.

\begin{figure}
\resizebox{7cm}{!}{\includegraphics{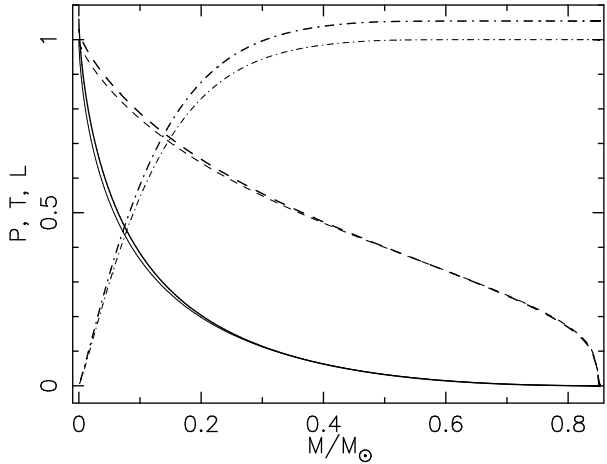}}
\caption{
The same, see Fig.~\ref{fig:abon}, for normalized pressure (full), temperature (dash)
and luminosity (dot-dash). The factors of normalization are respectively
$4.7\,10^{17}$ dyne cm$^{-2}$ for the pressure, $1.7\,10^7$ K for
the temperature and 
$3.6\,10^{33}$ erg cm$^{-2}$ s$^{-1}$ for the luminosity.
}\label{fig:struct}
\end{figure}

\section{Effects of microscopic diffusion on the structure of main-sequence
models}\label{sec:diff} 
Due to the gravitational settling, all species but \element[][1]{H}\ sink;
therefore as time goes on, the amount of hydrogen at the surface
is enhanced while
the amount of helium and heavy elements decreases. Then a first effect
of microscopic diffusion is the decrease of the surface value of [Fe/H] with
respect to time.
The variation of Z  modifies the opacity and then changes slightly
the temperature gradient in
the radiative zones, as well as the position of their boundaries.

In the mean time, the sinking of helium with respect to hydrogen produces a 
 decrease of the mean molecular weight
$\mu$ in the envelope as, assuming here a complete ionisation~:
\[
\mu\simeq\left(2{\rm X}+\frac34{\rm Y}+\frac {\rm Z}2\right)^{-1}
\,{\rm and}\ d\mu\simeq-\mu^2(\frac54dX+\frac{dZ}2).
\]
 In the outer layers of a model computed
with microscopic diffusion
$dX\geq0$, $dY\leq0$ and $dZ\leq0$ then $d\mu<0$ as $\mathrm{Z\ll X}$,
see Fig.~\ref{fig:abon}.

This variation of the mean molecular weight produces an increase of  the radius
 and consequently a decrease of the effective temperature.
But,  the variation of the metal content introduces a concommitant
variation of the opa\-ci\-ty in the radiative regions, which tend to
counterbalance this effect.

With the initial abundances given in Table~\ref{tab:x}, we have computed
respectively sets of models, i) without diffusion, 
ii) with diffusion of helium only, iii) with diffusion of all species.

The main
characteristics of four models of masses 0.85$M_\odot$, 0.8$M_\odot$ and
0.75$M_\odot$, at the age of 10\,Gyr computed without and with microscopic
diffusion of all species are displayed in Table~\ref{tab:a} and
in Table~\ref{tab:b} for the models of 0.7$M_\odot$ and
0.6$M_\odot$\footnote{There, for the calculations of [Fe/H]
with Eq. (\ref{eq:fesh}), the differential changes
between $\mathrm{Z_{Fe}}$ and Z, due to gravitational settling
and radiative forces, which act on opposite way, are neglected and 
$\mathrm{\log(Z_{Fe}/Z)\equiv-1.371}$ is used.}.

\begin{figure}
\resizebox{7cm}{!}{\includegraphics{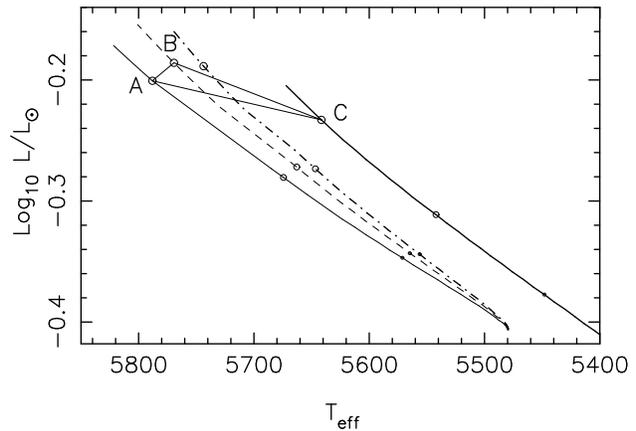}}
\caption{
The main-sequence HR evolutionary tracks for 0.8$M_\odot$, $\mathrm{[Fe/H]} =-0.94$,
without diffusion (thin full),
 with diffusion of helium only (dash-dot-dash), with diffusion of
 all elements (dashed), and
finally with the calibration of surface [Fe/H] (heavy full)
(the initial metallicity  has been computed in order to achieve, 
at 10\,Gyr, the initial surface metallicity
of the other three models). On each evolutionary track
the open circles correspond to ages of 3\,Gy, 6\,Gy and 9\,Gy respectively.
For 9\,Gy, AB is the diffusion shift {\bf DS},
BC the calibration shift  {\bf CS} and AC the global shift  {\bf GS}
(see text).
}\label{fig:dydc}
\end{figure}

During the main-sequence, models with diffusion
are cooler than models without diffusion; this is essentially due to the
changes of the mean molecular weight.
As already stressed by Castellani et al. (\cite{cas}) comparison with
unphysical models, including
diffusion of helium only, shows that a significant contribution
to the displacement in the HR diagram is 
due to helium settling, in particular for low metal content
as illustrated on Fig.~\ref{fig:dydc}.

In the following, we will call "diffusion shift" {\bf DS}, the
translation in the HR diagram
from a "standard model" to a model including microscopic diffusion,
at the {\em same mass and the same age}. It consists in~:
\begin{itemize}
\item a reduction of the effective temperature, decreasing
with mass from 100 K
at 0.85$M_\odot$, to almost zero at 0.6$M_\odot$; the stronger
 increase when [Fe/H] decreases is explained
by the larger influence of helium stratification at low metallicities;
\item an augmentation of the luminosity of the order of 0.01 
to 0.02 dex, almost independent of
[Fe/H], very sightly decreasing with mass, almost certainly due to 
the higher temperature
and higher helium content of the central core of models with diffusion.
\end{itemize}
The time dependence of the process shows up through the increasing
separation between the tracks as  evolution proceeds.

\begin{table*}
\caption[]{
Characteristic of models of $0.85M_\odot$ (Q), $0.80M_\odot$ (N), $0.75M_\odot$ (R),
$0.70M_\odot$ (P) and $0.60M_\odot$
(S) at 10Gy. The models evolved with
microscopic diffusion are specified by the labels 'dc' and 'd'
respectively for calibrated and non-calibrated (see text).
$M_{\rm bol}$ is the bolometric magnitude,
($M_{\odot\,\rm bol}\equiv4.75$); 
${\rm X}_\star$ and ${\rm Y}_\star$ are respectively
the surface abundances, in mass unit, of hydrogen and helium;
${\rm X}_0$ and ${\rm Z}_0$ are respectively
the initial surface abundances, in mass unit, of hydrogen and heavy
elements of calibrated models "dc";
$\mathrm{[Fe/H]}_0$ and $\mathrm{[Fe/H]}_\star$
are respectively the surface values
of [Fe/H] at zero age main-sequence and at 10Gy; $\mathrm{GS_T}$ in Kelvin,
is the component on the $T_{\rm eff}$ axis of the global shift
{\bf GS} (see text) for 10\,Gy. 
}\label{tab:a}
\begin{tabular}{lllllllllllllllll} \\  \hline \\
                       & Q07   & Q09   & Q12   & Q17   &  N07  &  N09  & N12   & N17   & R07   & R09   & R12   & R17   \\
\\ \hline \\
$T_{\rm eff}$          &5832   &6097   &6401   &6720   &5545   &5822   &6144   &6430   &5222   &5500   &5832   &6116   \\
$M_{\rm bol}$          &4.972  &4.649  &4.243  &3.811  &5.447  &5.179  &4.869  &4.573  &5.903  &5.664  &5.398  &5.163  \\
\\ \hline \\
                       & Q07d  & Q09d  & Q12d  & Q17d  & N07d  & N09d  & N12d  & N17d  & R07d  & R09d  & R12d  & R17d  \\
\\ \hline \\
$T_{\rm eff}$          &5809   &6060   &6322   &6584   &5531   &5801   &6101   &6363   &5215   &5488   &5810   &6080   \\
$M_{\rm bol}$          &4.925  &4.592  &4.174  &3.702  &5.408  &5.136  &4.818  &4.510  &5.869  &5.628  &5.358  &5.119  \\
${\rm X}_\star$        &0.802  &0.821  &0.853  &0.880  &0.795  &0.807  &0.825  &0.859  &0.789  &0.797  &0.808  &0.825  \\
${\rm Y}_\star$        &0.192  &0.177  &0.146  &0.120  &0.200  &0.190  &0.174  &0.141  &0.206  &0.200  &0.190  &0.175  \\
$\mathrm{[Fe/H]}_\star$&$-0.85$&$-1.07$&$-1.45$&$-2.00$&$-0.83$&$-1.04$&$-1.38$&$-1.94$&$-0.82$&$-1.02$&$-1.35$&$-1.86$\\
\\ \hline \\
                       & Q07dc & Q09dc & Q12dc & Q17dc & N07dc & N09dc & N12dc & N17dc & R07dc & R09dc & R12dc & R17dc \\
\\ \hline \\
${\rm X}_0$            &0.753  &0.755  &0.757  &0.759  &0.753  &0.756  &0.758  &0.759  &0.753  &0.756  &0.757  &0.759  \\
${\rm Z}_0$            &0.007  &0.005  &0.003  &0.001  &0.007  &0.005  &0.002  &0.001  &0.007  &0.004  &0.002  &0.001  \\
$T_{\rm eff}$          &5674   &5920   &6202   &6477   &5415   &5672   &5999   &6293   &5117   &5405   &5751   &6054   \\
$M_{\rm bol}$          &5.077  &4.779  &4.382  &3.892  &5.519  &5.262  &4.937  &4.590  &5.955  &5.698  &5.412  &5.144  \\
${\rm X}_\star$        &0.797  &0.811  &0.835  &0.872  &0.791  &0.801  &0.816  &0.849  &0.786  &0.795  &0.805  &0.822  \\
${\rm Y}_\star$        &0.197  &0.185  &0.163  &0.128  &0.203  &0.195  &0.182  &0.151  &0.208  &0.201  &0.193  &0.117  \\ 
$\mathrm{[Fe/H]}_0$    &$-0.66$&$-0.82$&$-1.08$&$-1.50$&$-0.67$&$-0.84$&$-1.12$&$-1.56$&$-0.68$&$-0.88$&$-1.17$&$-1.65$\\
$\mathrm{[Fe/H]}_\star$&$-0.75$&$-0.94$&$-1.25$&$-1.76$&$-0.75$&$-0.94$&$-1.24$&$-1.76$&$-0.75$&$-0.96$&$-1.27$&$-1.78$\\
$\mathrm{GS_T}$	       &$-158$ &$-177$ &$-199$ &$-243$ &$-130$ &$-150$ &$-145$ &$-137$ &$-105$ &$-95 $ &$-81 $ &$-62 $ \\
\\ \hline \\
\end{tabular}
\end{table*}

\begin{table*}
\caption[]{
Same as Table \ref{tab:a} for models of $0.70M_\odot$ (P) and $0.60M_\odot$ (S).
}\label{tab:b}
\begin{tabular}{lllllllllllllllll} \\  \hline \\
                       & P07   & P09   & P12   & P17   &  S07  &  S09  & S12   & S17   \\
\\ \hline \\
$T_{\rm eff}$          &4885   &5146   &5473   &5774   &4275   &4453   &4699   &4983   \\
$M_{\rm bol}$          &6.354  &6.137  &5.895  &5.688  &7.264  &7.087  &6.883  &6.701  \\
\\ \hline \\ 
                       & P07d  & P09d  & P12d  & P17d  &  S07d & S09d  & S12d  & S17d  \\
\\ \hline \\
$T_{\rm eff}$          &4880   &5139   &5461   &5756   &4277   &4453   &4699   &4983   \\
$M_{\rm bol}$          &6.325  &6.105  &5.861  &5.653  &7.239  &7.061  &6.856  &6.673  \\
${\rm X}_\star$        &0.784  &0.791  &0.798  &0.807  &0.776  &0.782  &0.786  &0.789  \\
${\rm Y}_\star$        &0.211  &0.206  &0.200  &0.193  &0.218  &0.215  &0.212  &0.211  \\
$\mathrm{[Fe/H]}_\star$&$-0.81$ &$-1.01$&$-1.32$&$-1.82$&$-0.79$&$-0.99$&$-1.30$&$-1.78$\\
\\ \hline \\
                       & P07dc & P09dc & P12dc & P17dc & S07dc & S09dc & S12dc & S17dc \\
\\ \hline \\
${\rm X}_0$            &0.754  &0.755  &0.758  &0.759  &0.753  &0.756  &0.758  &0.759  \\
${\rm Z}_0$            &0.006  &0.004  &0.002  &0.001  &0.006  &0.004  &0.002  &0.001  \\
$T_{\rm eff}$          &4803   &5072   &5413   &5742   &4239   &4420   &4672   &4979   \\
$M_{\rm bol}$          &6.394  &6.159  &5.898  &5.664  &7.280  &7.092  &6.876  &6.675  \\
${\rm X}_\star$        &0.782  &0.789  &0.797  &0.806  &0.775  &0.781  &0.785  &0.789  \\
${\rm Y}_\star$        &0.212  &0.207  &0.201  &0.194  &0.219  &0.216  &0.213  &0.211  \\
$\mathrm{[Fe/H]}_0$    &$-0.69$&$-0.89$&$-1.19$&$-1.68$&$-0.71$&$-0.90$&$-1.21$&$-1.71$\\
$\mathrm{[Fe/H]}_\star$&$-0.75$&$-0.96$&$-1.27$&$-1.78$&$-0.75$&$-0.95$&$-1.26$&$-1.77$\\
$\mathrm{GS_T}$	       &$-83 $ &$-74$  &$-60$  &$-32$  &$-36$  &$-33$  &$-27$  &$-4$  &\\
\\ \hline \\
\end{tabular}
\end{table*}

\section{"Calibration" of the subdwarfs main-sequence}\label{sec:cal}
Due to the decrease of the surface value of [Fe/H] along the evolution, the stars
presently observed with a given $\mathrm{[Fe/H]}_\star$ have started their
evolution with a larger $\mathrm{[Fe/H]}_0$ value of their metal content.

To represent observed data, models have then to be "calibrated" to
reach the observed surface $\mathrm{[Fe/H]}_\star$ value at their present age.
Assuming a linear dependence $\Delta\mathrm{[Fe/H]}/\mathrm{[Fe/H]}$
with respect to mass and age,
for a given mass  and a given age, one can estimate the initial
$\mathrm{[Fe/H]}_0$ needed to reach the observed value, and start again a new
evolutionary sequence 
including diffusion. If needed, this process can be iterated.

This effect adds a new shift of the HR
position of the models towards the red, called the "calibration shift" {\bf CS},
see Fig.~\ref{fig:dydc}. It also increases with age, as it is due to a
reduction of the surface abundances as evolution proceeds.
It consists in~:
\begin{itemize}
\item a decrease of the effective temperature due to the higher
initial metal content, which is higher for
higher metal content and decreases with mass from 140\,K
at 0.85$M_\odot$, to 30\,K
at 0.6$M_\odot$, for $\mathrm{[Fe/H]} = -0.94$.
\item a reduction of the luminosity, strongly depending on mass and becoming
very small for  0.6$M_\odot$, which is also due to the higher metal content.
\end{itemize}

The extend of {\bf CS} depends on the relative variation 
of $\mathrm{[Fe/H]}$ during the evolution, as seen from 
Table~\ref{tab:a} and \ref{tab:b} where are displayed the
main characteristics of
the models with enhanced initial
$\mathrm{[Fe/H]}_0$ (label "dc") in order to obtain at the end of
the evolution, the surface values of
$\mathrm{[Fe/H]}_\star$ close to the values of Table~\ref{tab:x}.

\begin{figure}
\resizebox{7cm}{!}{\includegraphics{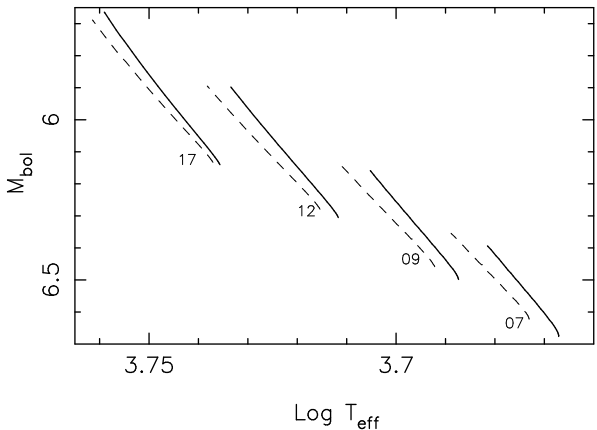}}
\caption{
The main-sequence evo\-lu\-tion\-ary tracks of $0.7\,M_\odot$ models,
evolved without diffusion (dashed) and calibrated (full).
The solar bolometric magnitude is taken as $M_{\odot\,\rm bol}\equiv4.75$.
}\label{fig:modp}
\end{figure}

\begin{figure}
\resizebox{7cm}{!}{\includegraphics{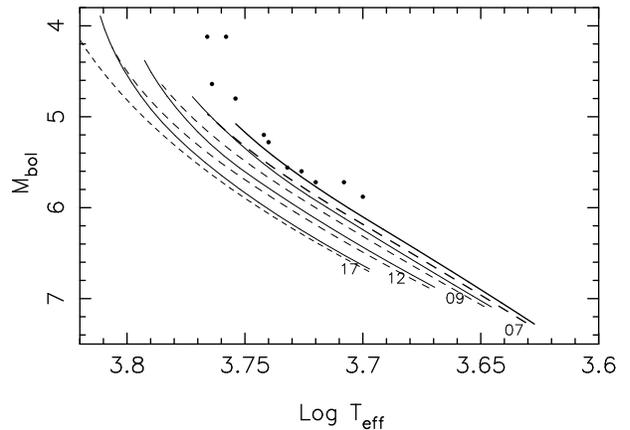}}
\caption{
Schematic isochrones at 10\,Gy for models with
$\mathrm{[Fe/H]}_\star=-0.75$, $-0.94$,
$-1.24$ and $-1.72$; dashed~: model without diffusion, full~:
calibrated models with diffusion. $\bullet$~: locations of stars
with observed [Fe/H]
within $\mathrm{-1.0<[Fe/H]<-0.45}$ as derived from Hipparcos data.
}\label{fig:iso10}
\end{figure}

\begin{figure}
\resizebox{7cm}{!}{\includegraphics{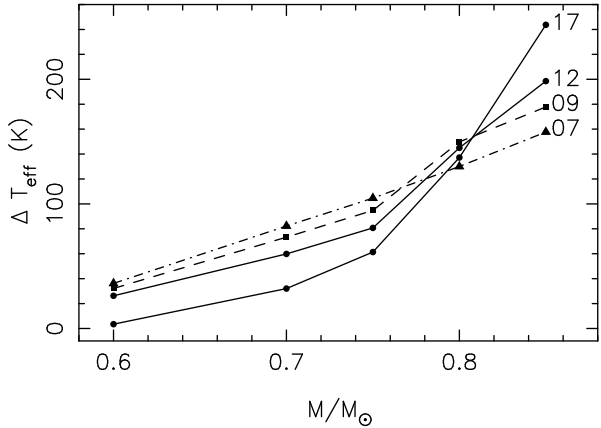}}
\caption{
Total variation of effective temperature at 10\,Gy
between standard and calibrated models
with respect to the mass. At  $0.85\,M_\odot$
evolutionary effects become significant.
}\label{fig:dtdm}
\end{figure}

\begin{figure}
\resizebox{7cm}{!}{\includegraphics{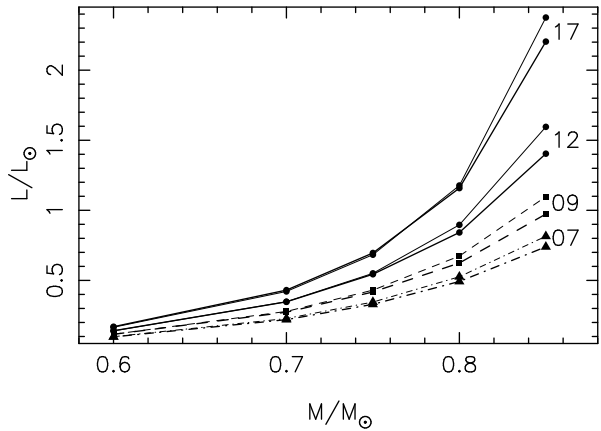}}
\caption{Mass-luminosity diagram at 10\,Gy for standard (thin)
and calibrated (thick) models.
}\label{fig:lm}
\end{figure}

\section{ The global shift}\label{sec:HR}
We will call "global shift" {\bf GS}, 
the combined effect of diffusion and calibration. 
Its component in effective temperature is
 given in Table~\ref{tab:a} and \ref{tab:b}
for the  10\,Gy models. Fig.~\ref{fig:dtdm} 
shows, as a function of mass, the variations of effective temperature 
for the three different metal content considered here.

Both diffusion and calibration reduce the effective temperature, but in an
opposite way with respect to
[Fe/H]. The effect of diffusion increases while the effect of
calibration decreases with [Fe/H].

\begin{figure}
\resizebox{7cm}{!}{\includegraphics{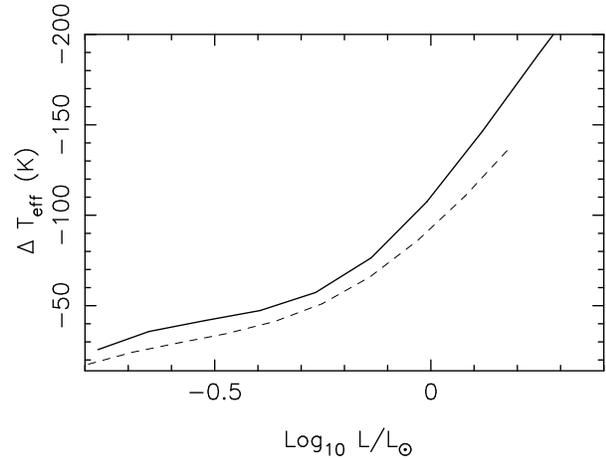}}
\caption{
Temperature difference between two isochrones at the same
luminosity as a function of luminosity, respectively
for 10\,Gy (full) and 8\,Gy (dahed), with $\mathrm{[Fe/H]}_\star=-1.72$
-- wigles are due to the interpolation procedure.
}\label{fig:dteff15}
\end{figure}

In luminosity the situation is different. The calibration procedure 
compensates partly the luminosity
increase due to diffusion. The variation can reach 0.03 dex, depending 
both on mass and [Fe/H]. The mass dependence remains strong,
from 0.06\,dex at 0.85$M_\odot$, to
0.01\,dex at 0.6$M_\odot$.
It produces a distortion in the Mass-Luminosity relation, see Fig.~\ref{fig:lm},
depending on [Fe/H], an effect which
could be detected with high quality data on a reasonable set of objects.

The displacement in the HR diagram is seen as a translation towards
lower effective temperatures, see Fig.~\ref{fig:modp}. It produces
 a slight variation of
the curvature of the evolutionary track, see Fig.~\ref{fig:iso10}, due to
the time dependence of the process. Isochrones are then slightly more vertical.

Stars with observed [Fe/H]
within $\mathrm{-1.0 < [Fe/H] < -0.45}$ and the most accurate
fundamental parameters, studied by 
 Lebreton et al. (\cite{lpf}) have been added to Fig.~\ref{fig:iso10}. As
already said, though  microscopic diffusion tends to reduce the discrepancy
between the effective temperature of observed and theoretical models, the
effect  is 
 not sufficient to solve the difficulty.

For $\mathrm{[Fe/H]} =-1.72$, Fig.~\ref{fig:dteff15} shows, as a
function of luminosity and effective temperature,
 the effective temperature difference
between two isochrones at the same luminosity~:
the standard one
experiences no diffusion and its initial abundances are equal to the observed ones,
while the second one is computed using our proposed calibration.

\section{Discussion}\label{sec:fut}
We have estimated in this paper the influence of gravitational and
thermal settling, on the global parameters of low main-sequence stars.
As we have introduced only microscopic flows
in the computations except in convective zones, the stratification is
 probably overestimated.

We have shown that this physical process, if at work permanently,
produces an effective
temperature shift towards the red. For population II stars
it exceeds generally 100 K.

Compared to the most reliable data available at present, 
the predicted shift is of the right order of magnitude,
but probably slighly unsufficient
to eliminate completely the discrepancy shown by Lebreton et al.
(\cite{lpf}).
However, as already mentioned, other effects could also be responsible 
for redshifts of that kind. 
So that more precise characteristic behaviors are needed to decipher 
among the different possibilities.
Let us cite a few signatures of the microscopic diffusion.
\begin{itemize}
\item the predicted shift increases with age and
decreases when [Fe/H]  increases;
\item the mass-luminosity relationship is modified depending
on Z as at a given mass
and age, as diffused and calibrated models are  less luminous than standard ones;
\item the slope of isochrones is slightly increased.
\end{itemize}
Though very important improvement on the quality of the
observational data on the nearest members of the old
population have been obtained recently,
they remain too coarse and too few to allow unambiguous tests.
More objects with accurate fundamental parameters are needed
to validate definitely this
hypothesis.

A lot of progress is expected from the GAIA mission
(Perryman et al. \cite{pergaia}),
which will provide even more accurate distances and sometimes masses for a larger
number of stars. But improvements in the
determination of atmospheric parameters, as well as
in methods of comparing observed and
predicted quantities are also badly needed.

\begin{acknowledgements}
We are grateful to G. Alecian, Y. Lebreton and R. Cayrel for helpful
discussions and communicating their work before publication.
We would also like to thank the referees,
Drs. S. Degl'Innocenti and Dr. V. Castellani, for their careful
review of this paper and for a number and useful suggestions.
This work was partly supported by the GDR G131 "Structure
Interne des Etoiles et des Plan\`etes g\'eantes" of CNRS (France).
Most of this work has been performed using the computing facilities 
provided by the OCA program
``Simulations Interactives et Visualisation en Astronomie et M\'ecanique 
(SIVAM)''.
\end{acknowledgements}

\end{document}